\documentclass[12pt]{article}

\usepackage{tabularx} %Special tables
\usepackage{rotating} % rotate Boxes
\usepackage{epsfig}

\usepackage{multirow} %Create multirow entries on a table
\usepackage{graphicx} %Graphics
\usepackage{pstricks}
\usepackage{amssymb}

\newcommand{\gsim}{\mbox{${~\raise.25em\hbox{$>$}\kern-.70em
\lower.25em\hbox{$\sim$}~}$}}
\newcommand{\lsim}{\mbox{${~\raise.25em\hbox{$<$}\kern-.70em
\lower.25em\hbox{$\sim$}~}$}}

\author{
E. Nardi$^{1,2}$,  Y. Nir$^3$, J. Racker$^4$ and E. Roulet$^4$  \\[5pt]
$^1${\normalsize \it INFN, Laboratori Nazionali di Frascati, C.P. 13, 
I00044 Frascati,  Italy}  \\[-2pt]
$^2${\normalsize \it Instituto de F\'\i sica,  Universidad de Antioquia, 
A.A.{\it 1226}, Medell\'\i n, Colombia} \\ [-2pt]
$^3${\normalsize \it Department of Particle Physics, Weizmann Institute of  
Science,  Rehovot 76100, Israel}\\[-2pt]
$^4${\normalsize \it CONICET, Centro At\'omico Bariloche,  Avenida Bustillo
  9500 (8400) Argentina}}

\title{\vspace{-3cm} On Higgs and sphaleron effects
during the leptogenesis era}

\begin{document}

\maketitle

\begin{abstract}
We discuss the effects of various processes that can be active during
the leptogenesis era, and present the Boltzmann equations that take
them into account appropriately. A non-vanishing Higgs number
asymmetry is always present, enhancing the washout of the lepton
asymmetry. This is the main new effect when leptogenesis takes place
at $T>10^{12}$~GeV, reducing the final baryon asymmetry and tightening
the leptogenesis bound on the neutrino masses. If leptogenesis occurs
at lower temperatures, electroweak sphalerons partially transfer the
lepton asymmetry to a baryonic one, while Yukawa interactions and QCD
sphalerons partially transfer the asymmetries of the left-handed
fields to the right-handed ones, suppressing the washout processes.
Depending on the specific temperature range in which leptogenesis
occurs, the final baryon asymmetry can be enhanced or suppressed by
factors of order 20\%--40\% with respect to the case when these effects
are altogether ignored.
\end{abstract}

%%%%%%%%%%%%%%%%%%%%%%%
\section{Introduction}
One of the most attractive scenarios to explain the origin of the baryon
asymmetry of the Universe ($Y_B\equiv (n_B-{\bar n}_B)/s\simeq 8.7 \times
10^{-11}$) is leptogenesis \cite{fu86,lu92}. In this framework the decays of
heavy electroweak singlet neutrinos (such as those appearing in see-saw
models) into lepton and Higgs particles generate a lepton asymmetry, which is
then partially reprocessed into a baryon asymmetry by anomalous electroweak
processes mediated by sphalerons.

To compute in detail the lepton (and baryon) asymmetry at the end of the
leptogenesis era, one has to take into account the various processes which
can modify the particle densities. Some of these, such as the heavy neutrino
decays or the various interactions that can washout the lepton number,
occur slowly as compared to the expansion rate of the Universe, and hence are
naturally accounted for via appropriate Boltzmann equations. Other
reactions can be very fast (depending on the temperature considered)
and their effect is to impose certain relations among the chemical
potentials of different particle species that hold within specific
temperature ranges. These include the Standard Model gauge
interactions, some Yukawa interactions involving heavy fermions,
and electroweak and strong non-perturbative `sphaleron' processes.
In the present work we analyze how all these ingredients concur to
determine the precise impact of the washout processes and we discuss
the effects this has on the final value of the baryon asymmetry.

A final important phenomenon has to do with the flavor composition of
the states involved \cite{ba00}, and the decoherence effects induced
by the leptonic Yukawa interactions in equilibrium, which essentially
act as measuring devices that project the densities onto the flavor
basis. The flavor interplay between the lepton number violating
processes and the Yukawa interactions is very rich and can lead to
dramatic consequences. Here, for the sake of clarity, we assume a
simple flavor structure. We will discuss the full flavor picture in a
separate publication \cite{future}.

The main point where our paper provides new insights is the combined
effect of all spectator processes -- QCD sphalerons, electroweak sphalerons and
Yukawa interactions --  on the washout processes for the various
relevant temperature regimes. In particular, we emphasize the role of the Higgs
number asymmetry. The issue of Higgs processes has been raised and analyzed in
ref. \cite{bu01}. We improve upon their analysis at several points
and, in the end, are led to opposite conclusions regarding the
direction of the effect, at least for some temperature regimes.
Spectator processes in the low temperature regime (region 6 of
section 4.2 below) were appropriately taken into account in ref.~\cite{pi05}.

The plan of this paper is as follows. In Section \ref{sec:basic} we
present our framework and we enumerate and parametrize the various
relevant washout processes. In Section \ref{sec:boltzmann} we discuss
the Boltzmann equations. In Section \ref{sec:equi} we present our main
results. Equilibrium conditions in the various temperature regimes are
imposed, and the implications for the Boltzmann equations are
analyzed. Results are obtained for various representative
flavor-alignment structures. In Section \ref{sec:mbar} we explain
how the leptogenesis bound on the absolute scale of neutrino masses
is affected by our considerations. 

%%%%%%%%%%%%%%%%%
\section{The Basic Framework}
\label{sec:basic}
We consider for simplicity the scenario in which right
handed neutrino masses are hierarchical, $M_1\ll M_{2,3}$, and
consequently the lepton asymmetry is mainly generated via the CP and lepton
number violating decays of the lightest singlet neutrino
$N_1$. Even in this case, the task of including a general flavor
structure within the Boltzmann equations can be quite complicated.  In
the mass eigenstate basis of the heavy neutrinos $N_\alpha$
($\alpha=1,2,3$) and of the charged leptons ($i=e,\mu,\tau$), the
Yukawa interactions for the leptons read
\begin{equation}
\label{yuk} 
{\cal L}_Y=
-h_{i \alpha}\,\overline{N}_\alpha \ell_i\,  {\widetilde H}^\dagger
-h_{i}\, \overline{e}_i \ell_i\, H^\dagger
+h.c.,
\end{equation}
where $\ell_i$ and $e_i$ denote the $SU(2)$ lepton doublets and singlets,
$H=(H^+,H^0)^T$ is the Higgs field ($\widetilde H =i\tau_2 H^*$) and the
couplings $h$ for $e_i$ and $N_\alpha$ can be easily distinguished by the
presence of one or two indices.  

It is convenient to define a lepton doublet state $\ell_D$ as follows:
\begin{equation}
  \ell_D = \frac{h_{i1 }}{\sqrt{(h^\dagger h)_{11}}}\,\ell_i.
\label{eqelld}
\end{equation} 
The state $\ell_D$ is the one appearing (at tree level) 
in the following relevant processes:
\begin{itemize} \itemsep=1pt
\item $N_1$ decays and inverse decays, with rate $\gamma_D=\gamma(N_1
  \leftrightarrow \ell_D\, H)$;
\item $\Delta L=1$ Higgs-mediated scattering processes with rates such
as $\gamma_{S_s} 
=\gamma(\ell_D\, N_1 \leftrightarrow Q_3\, t)$ and $\gamma_{S_t}
=\gamma(\ell_D\, Q_3 \leftrightarrow N_1\, t)$ , where $Q_3$ and $t$
are respectively the third generation quark doublet and the top
$SU(2)$ singlet, as well as those involving gauge bosons, such as in
$\ell_D N_1\to HA$  (with $A=W_i $ or $B$).
\item The on-shell $N_1$-mediated $\Delta L=2$ scattering processes
contributing to the rate $\gamma_{Ns} = \gamma(\ell_D\, H
\leftrightarrow \overline{\ell}_D\, \overline{H})$.
\end{itemize}
In terms of $\ell_D$, the neutrino Yukawa interaction of $N_1$ in
eq.~(\ref{yuk}) reads $-\sqrt{(h^\dagger h)_{11}}\,\overline{N}_1
\ell_D\,  {\widetilde H}^\dagger$. Consequently, all the rates above
depend on a single combination of 
neutrino Yukawa couplings, and are often parametrized
in terms of a dimensional parameter $\tilde m_1$:
\begin{equation}\label{tildem}
  \tilde m_1\equiv (h^\dagger h)_{11}\frac{v^2}{M_1},
\end{equation}
where $v=\langle H\rangle$.
There is one additional class of relevant lepton number violating processes:
\begin{itemize}
\item Off-shell contributions to $\Delta L=2$ scattering
processes modify $\gamma_{Ns}$ and induce $\gamma_{Nt} =
\gamma(\ell\, \ell \leftrightarrow \overline{H}\, \overline{H})$.
\end{itemize}
These processes are mediated by heavy neutrino exchanges, the first in
the $s$- and $u$-channels while the second in the $t$-channel.  The
amplitude for the off-shell contributions to $\Delta L=2$ washout is
essentially proportional, in the limit $T<M_1$, to the light neutrino
mass matrix, ${\cal M}_{ij} = h_{i\alpha }\frac{v^2}{M_\alpha}
h_{j\alpha }\,$. Consequently, the fastest rate couples to the lepton
doublet containing the heaviest light neutrino state $\nu_3$.  We see
then that in general it is not only the state $\ell_D$ which is
involved in these contributions, further complicating the flavor
structure of the problem.

If the resonant contribution to the scattering is properly subtracted
\cite{gi04}, one finds that these off-shell pieces are generally
sub-dominant, so that the $\Delta L=2$ washout processes are in general
dominated by the on-shell $N_1$ exchange (with a possible exception if
$M_1 \gg 10^{12}$~GeV, in which case some Yukawa couplings can become
of order unity).  In this case, the only direction in flavor space
that appears in the lepton number violating processes is that of
$\ell_D$\footnote{Actually, the state $\ell_D$ produced  in $N_1$
 decays differs from the one in eq.~(\ref{eqelld}) at
one-loop \cite{ba00}.
 Moreover, at one loop level
the anti-lepton produced in $N_1$ decays $\bar\ell_D$
is not necessarily the conjugate of $\ell_D$, and this
can have interesting effects, which will be explored in \cite{future}.}. 
In the following, we make the assumption that this is indeed
the case and we often use the simplified notation $\ell\equiv\ell_D$
for this special direction. This assumption simplifies things
considerably, because otherwise it becomes necessary to follow the
evolution of the asymmetries in an approach based on Boltzmann
equations for the density matrix \cite{ba00} and keeping track of
coherence effects. 

Another issue related with flavor becomes quite important at
temperatures when the processes induced by the Yukawa couplings of the
charged leptons, $h_{\tau,\mu,e}$ of eq.~(\ref{yuk}), are no more
negligible. The lepton Yukawa interactions define a flavor basis, and
the density matrix for the lepton asymmetry is projected onto this
basis.  If the state $\ell$ is not aligned with a specific flavor
$\ell_\tau$, $\ell_\mu$ or $\ell_e$ then the lepton asymmetry gets
distributed between all the different flavors.
Such misalignment has many interesting consequences, which we will
present in \cite{future}.

%%%%%%%%%%%%%%%%%%%%%%
\section{The Boltzmann equations}
\label{sec:boltzmann}
In this section we present the Boltzmann equations that are relevant to
the washout effects in leptogenesis, focusing on the case when $\ell$ is
aligned along one specific flavor direction. This case can be treated
more easily and allows us to understand in detail the
flavor-independent effects that we want to explore in this paper.

With the simplifying alignment conditions discussed above, the (linearized) 
Boltzmann equations can be written  as:
\begin{eqnarray}
\frac{{\rm d}Y_N}{{\rm d}z}
&=&-\frac{1}{sHz}\left(\frac{Y_N}{Y_N^{eq}}-1\right)
\left(\gamma_D+2\gamma_{Ss}+4\gamma_{St}\right),
\label{eqyn} \\ [5pt]
\frac{{\rm d}Y_L}{{\rm d}z}&=&\frac{1}{sHz}\left\{
  \epsilon\left(\frac{Y_N}{Y_N^{eq}}-1\right)\gamma_D 
  - \left[ 2y_\ell+ 
    \left(y_t-y_{Q_3}\right)
    \left( \frac{Y_N}{Y_N^{eq}}+1\right) \right]\gamma_{St}
\right.\nonumber \\[5pt]
&-& \!\!\! \left.   
  \left( \frac{Y_N}{Y_N^{eq}}y_\ell
    +y_t-y_{Q_3} \right)\gamma_{Ss}
  - 2\left(y_\ell +y_H\right)
  \left(\gamma_{Ns}+\gamma_{Nt}\right)
\right\}+\frac{{\rm d}Y_L^{EW}}{{\rm d}z}\,,
\label{eqyl}    \\ [5pt]
\frac{{\rm d}Y_B}{{\rm d}z}&=&\frac{{\rm d}Y_B^{EW}}{{\rm d}z}\,, 
\label{eqyb} 
\end{eqnarray} 
where the standard notation $z\equiv M_1/T$ is used. Here $Y_N\equiv
n_N/s$ is the density of the lightest heavy neutrinos (with two
degrees of freedom) relative to the entropy $s$, $Y_L$ and $Y_B$ are
the total lepton and baryon number densities, also normalized to the
entropy, $y_X\equiv (n_X-n_{\bar X})/n_X^{eq}$ denote the asymmetries
for the relevant different species $X= \ell,\ H,\ t,\ Q_3\,$ and all
the asymmetries are normalized to the Maxwell-Boltzmann equilibrium
densities. The reaction rates are summed over initial and final state
quantum numbers, including the gauge multiplicities. In the
asymmetries $y_X$, $X=\ell$, $H$ or $Q_3$ label any of the two doublet
components, not their sum, and hence we normalize $y_X$ to the
equilibrium densities with just one degree of freedom. This is
different from the convention in \cite{gi04}, and allows us to keep
the proportionality $y_X\propto \mu_X$ in terms of the chemical
potentials, with the usual convention that {\it e.g.}  $\mu_{\ell_i}$
is the chemical potential of each one of the two $SU(2)$ components of
the doublet $\ell_i$.

In our analysis, we make two further simplifications:
\begin{enumerate}
\item In Eqs.~(\ref{eqyn})--(\ref{eqyb}) and in what follows we ignore
finite temperature corrections to the particle masses and couplings
\cite{gi04}, so that we take all equilibrium number densities $n_X^{eq}$
equal to those of massless particles.
\item We ignore scatterings involving gauge bosons, for whose rates no
consensus has been achieved so far \cite{gi04,pi04}. They do not
introduce qualitatively new effects and, moreover, no further density
asymmetries are associated to them.
\end{enumerate}

We would like to emphasize the following points concerning
eq.~(\ref{eqyl}): 
\begin{itemize}
\item The CP violating parameter $\epsilon$ gives a measure of the $L$
asymmetry produced per $N_1$ decay \cite{co96}. 
\item $y_{\ell}$ is the asymmetry for one component of the relevant 
$SU(2)$-doublet $\ell\equiv\ell_D$ (relative to equilibrium density),
while the total lepton asymmetry is $ Y_L= \sum_i Y_{L_i}=
\sum_i(2\,y_{\ell_i}+y_{e_i})Y^{eq}$, with $Y^{eq}\equiv n^{eq}/s$.
\item For the temperature regimes in which the charged lepton Yukawa
couplings become non-negligible ($T \lsim 10^{13}\,$GeV), the
corresponding interactions define a lepton flavor basis. We assumed,  
for simplicity, that the state $\ell$ is aligned with (or orthogonal 
to) one of the lepton flavor states singled out by the Yukawa
interactions. Then the difference in the rates $\Gamma\equiv
\Gamma(N_1\to \ell\, h)$ and $\bar \Gamma \equiv \Gamma(N_1\to \bar
\ell\, \bar h)$ for the $N_1$ decays into $\ell$ leptons and
$\bar\ell$ antileptons gives the total $CP$-asymmetry $\epsilon=
(\Gamma-\bar\Gamma)/(\Gamma+\bar \Gamma)$, while the evolution of
total lepton number is determined by the Boltzmann equation
(\ref{eqyl}) solely in terms of one leptonic asymmetry $y_\ell$.
However, in the general case of no alignments, the decay rates of
$N_1$ into the specific flavors $\ell_i$ and anti-flavors $\bar
\ell_i$ have to be considered, and the Boltzmann equations should
track the evolution of all the relevant single-flavor asymmetries
\cite{ba00,future}. 
\item The thermally averaged reaction rate $\gamma_{Ns}$ is the
$\Delta L=2$ $s$-channel rate without subtraction of the real
intermediate state, and thus it takes into account also the on-shell
heavy neutrino exchanges \cite{gi04}. Since we consider here only the
tree level scatterings, there is no double counting of the CP
violating one loop contribution included in the direct and inverse
decay terms.   
\item The factor d$Y_L^{EW}/{\rm d}z$ is included to account for
the lepton number violation induced by electroweak anomalous
processes. This term is proportional to the electroweak sphaleron rate
$\Gamma_{EW}$ and to the amount of $(B+L)$ asymmetry contained in the
left-handed fields. It becomes relevant at temperatures below
$10^{12}$~GeV. Since electroweak sphalerons preserve $B-L$, one has
\begin{equation}
\frac{{\rm d}Y_B^{EW}}{{\rm d}z}=
\frac{{\rm d}Y_L^{EW}}{{\rm d}z}.
\end{equation}
Hence, by subtracting eqs. (\ref{eqyl}) from (\ref{eqyb}), one can
combine them into a single equation for $Y_{B-L}$. The resulting
equation takes into account all the relevant washout processes, and
has the advantage of being independent of the (poorly known) sphaleron
rate. Note the the electroweak sphalerons preserve not only $B-L$, but
also the three lepton-flavor related quantities
\begin{equation}
  \Delta_i\equiv B/3-L_i.
  \label{defDeltai}
  \end{equation}
\end{itemize}

Equation (\ref{eqyl}) takes into account the fact that the heavy
neutrino decays, besides producing an asymmetry in the left-handed
leptons, also produce an asymmetry in the Higgs number density. The
Higgs number is not conserved by Yukawa interactions, but its
asymmetry is only partially transferred into a `chiral' asymmetry
between $Q_{3}$ and $t$ by the top quark Yukawa interactions (as well
as into asymmetries for other fermions when their corresponding
interactions with the Higgs enter into equilibrium). Indeed, the
equilibrium conditions enforce $y_H\neq 0$, and hence $y_H$ acts as a
source of washout processes.  Similarly, $y_t-y_{Q_3}$ acts as a
source for the $\Delta L=1$ washout processes involving Higgs boson
exchanges (washout processes involving standard model Yukawa couplings
different from the top one can be safely neglected). These additional
contributions to the washout of lepton number are often ignored in the
literature and omitted from the Boltzmann equations for leptogenesis,
yet they play a role that is similar, qualitatively and
quantitatively, to the role of $y_\ell$.\footnote{The additional
washout terms in eq.~(\ref{eqyl}) were considered before in
\cite{bu01}. We improve this analysis by giving a proper treatment of
the $B-L$ conserving electroweak sphaleron processes, by 
coupling the washout terms only to the relevant lepton doublet
asymmetry, and by accounting for the QCD sphalerons as well as for all
the other processes that enter into equilibrium at the different
temperature regimes. This leads to results that differ  from those
of ref. \cite{bu01}.}

The system of equations that now has to be solved corresponds to
eq.~(\ref{eqyn}) for $Y_N$ and the equation derived from subtracting
eq.~(\ref{eqyl}) from eq.~(\ref{eqyb}) for $Y_{B-L}$. This system can
be solved after expressing the densities $y_{\ell}$,  $y_H$ and
$y_t-y_{Q_3}$ in terms of $Y_{B-L}$ with the help of the equilibrium
conditions imposed by the fast reactions, as described in the next
section. The value of $B-L$ at the end of the leptogenesis era
obtained by solving the Boltzmann equations remains subsequently
unaffected until the present epoch. If electroweak sphalerons go out
of equilibrium before the electroweak phase transition, the present
baryon asymmetry (assuming a single Higgs doublet) is given by the
relation~\cite{ha90} 
\begin{equation}
n_B=\frac{28}{79}n_{B-L}.
\label{bvsbml}
\end{equation}
If, instead, electroweak sphalerons remain in equilibrium until
slightly after the electroweak phase transition (as would be the case
if, as presently believed, the electroweak phase transition was not
strongly first order) the final relation between $B$ and $B-L$ would
be somewhat different \cite{la00}. 

%%%%%%%%%%%%%%%%%%%%%%%%%%%%%%%%%%%%%%%%%%%%%%%%%%%%
\section{The equilibrium conditions}
\label{sec:equi}
In this section we  discuss the equilibrium conditions that hold in the
different temperature regimes which can be relevant for leptogenesis.
Since leptogenesis takes place during the out of equilibrium decay of the
lightest heavy right-handed neutrino $N_1$, {\it i.e.} at temperatures
$T\sim M_1$, the relevant constraints that have to be imposed among
the different asymmetries depend essentially on the value of $M_1$. We
use the equilibrium conditions to express $y_\ell$, 
$y_H$ and $y_t-y_{Q_3}$ in
terms of $Y_{B-L}$.  

%%%%%%%%%%%%%%%
\subsection{General considerations}
The number density asymmetries for the particles $n_X$ entering in
eq.~(\ref{eqyl}) are related to the corresponding chemical potentials
through 
\begin{equation}
n_X-n_{\bar X}=
\frac{g_XT^3}{6}\cases{\mu_X/T&fermions,\cr 2\mu_X/T&bosons,\cr} 
\label{nvsmu.eq}
\end{equation}
where $g_X$ is the number of degrees of freedom of $X$.  For
any given temperature regime the specific set of reactions that are in
chemical equilibrium enforce algebraic relations between different chemical
potentials \cite{ha90}.
In the entire range of temperatures relevant for leptogenesis, the
interactions mediated by the top-quark Yukawa coupling $h_t$, and by
the $ SU(3)\times SU(2)\times U(1)$ gauge interactions, are always in
equilibrium. This situation has the following consequences:
\begin{itemize}
\item Equilibration of the chemical potentials for the
different quark colors is guaranteed because the chemical potentials
of the gluons vanish, $\mu_g=0$.
\item Equilibration of the chemical potentials for the two members of
an $SU(2)$ doublet is guaranteed by 
the vanishing, above the electroweak phase transition, of
$\mu_{W^+}=-\mu_{W^-}=0$. This condition was implicitly implemented
in eq.~(\ref{eqyl}) where we used $\mu_Q\equiv \mu_{u_L}=\mu_{d_L}$,\
$\mu_\ell\equiv \mu_{e_L}=\mu_{\nu_L}$ and $\mu_H=\mu_{H^+}=\mu_{H^0}$ to
write the particle number asymmetries directly in terms of the number
densities of the $SU(2)$ doublets.
\item Hypercharge neutrality implies
\begin{equation} \label{hyper}
\sum_i\left( \mu_{Q_i}+2\mu_{u_i}-\mu_{d_i}-\mu_{\ell_i}-\mu_{e_i}\right)+ 
2\mu_H =0\,, 
\end{equation}
where $u_i$, $d_i$ and $e_i$ denote the $SU(2)$ singlet fermions of
the $i$-th generation.
\item The equilibrium condition for the Yukawa interactions
of the top-quark $\mu_t = \mu_{Q_3}+\mu_H$ yields:
\begin{equation}
y_t-y_{Q_3}=\frac{y_H}{2}\,,
\end{equation}
where the factor 1/2 arises from the relative factor of 2 between the number
asymmetry and chemical potential for the bosons, see
eq.~(\ref{nvsmu.eq}).
\end{itemize}

Using this relation, one can recast the Boltzmann equation for the
$B-L$ asymmetry in the aligned case as
\begin{eqnarray}
  \frac{{\rm d}Y_{B-L}}{{\rm d}z}&=&  \frac{-1}{sHz}\left\{
    \left(\frac{Y_N}{Y_N^{eq}}-1\right)\,\left[\epsilon\, \gamma_D 
    +  \left(c_\ell\, \gamma_{Ss} + 
\frac{c_H}{2}\, \gamma_{St}\right)\frac{Y_{B-L}}{Y^{eq}}\right]
    \right. + \cr
  && 
  \hspace{-2cm} \left.  \phantom{\frac{Y_N}{Y^eq_N}}
\left[\left(2\,c_\ell + c_H\,\right)
\left(\gamma_{St}+\frac{1}{2}\gamma_{Ss}\right) + 2\left(c_\ell +
c_H\right)
\left(\gamma_{Ns}+\gamma_{Nt}\right)\right]\frac{Y_{B-L}}{Y^{eq}}\right\},
\label{eqybml2}    
\end{eqnarray}
where we have defined $c_\ell$ and $c_H$ through $ y_\ell \equiv
-c_{\ell}\, Y_{B-L}/Y^{eq}$ and $y_{H}\equiv -c_{H}\, Y_{B-L}/Y^{eq}$
while their numerical values are determined, within each temperature
range, by the constraints enforced by the fast reactions that are in
equilibrium.  This equation is general enough to account for all the
effects of the relevant spectator processes (Yukawa interactions,
electroweak and QCD sphalerons), while to take into account the lepton
flavor structure, a generalization of eq.~(\ref{eqybml2}) is required.

%%%%%%%%%%%%%%%
\subsection{Specific temperature ranges and flavor structures}
\label{subs:ranges}
Let us now discuss the different temperature ranges of interest. At
each step, we take into account all the relevant processes that enter
into equilibrium. In order to understand and disentangle the various
effects involved, we examine a rather large number of temperature
windows, and for each window we also impose, when relevant, various
conditions of flavor alignment.

Our main results can be understood on the basis of the examples
presented in Table~\ref{tab:results}, that cover six different
temperature regimes. For each regime, different possibilities of
flavor alignments are considered. To do that, we define a parameter
$K_i$ ($i=e,\mu,\tau$): 
\begin{equation}
  K_i\equiv |\langle\ell_i|\ell_D\rangle|^2.
  \label{definek}
\end{equation}
The simple flavor structures that we investigate here correspond to
either alignment with a specific flavor direction, $K_i=1$, or
orthogonality, $K_i=0$. The more general case of $K_i\neq0,1$ will be
discussed in \cite{future}.

For each aligned case, we give in the table the
coefficients $c_\ell$ and $c_H$ that relate the asymmetries $y_\ell$
and $y_H$ to $Y_{B-L}$. Note that $c_\ell$ and $c_H$ give a
crude understanding of the impact of the respective asymmetries:
$c_H/c_\ell$ gives a rough estimate of the relative contribution of
the Higgs to the washout, while $c_\ell + c_H$ gives a measure of the
overall strength of the washout.
The last column gives the resulting $B-L$ asymmetry for each case. To
disentangle the impact of the various processes from that of the input
parameters, the $B-L$ asymmetry is calculated in all cases with fixed
values of $\tilde m_1=0.06$~eV and $M_1=10^{11}$~GeV. The $\tilde m_1$
parameter was defined in eq.~(\ref{tildem}). It determines the
departure from equilibrium of the heavy neutrino $N_1$, as well as the
strength of the washout processes. For $\tilde{m}_1<10^{-3}$~eV, the
departure from equilibrium is large and washout effects are generally
negligible. Hence, in this case, there is no need to solve any
detailed Boltzmann equations. In contrast, for $\tilde{m}_1 \gsim 
0.1$~eV, washout processes become so efficient that, in general, the
surviving baryon asymmetry is too small. We therefore consider the
intermediate value  $\tilde{m}_1=0.06$~eV, which is also suggested
by the atmospheric neutrino mass-squared difference if neutrino masses
are hierarchical.  As concerns $M_1$, it is clear that the relevant
temperature range is actually determined by it, yet -- as explained
above -- we fix the value at $M_1=10^{11}$~GeV in order to have a
meaningful comparison of the various effects of interest. Namely,
since in each regime considered the same asymmetries are produced in the
decay of the heavy neutrinos, a comparison between the final values of
$B-L$ for the different cases can be directly interpreted in terms of
suppressions or enhancements of the washout processes. 
Anyhow, the overall effects of the washouts turn out to be
essentially independent of the values of $M_1$, as long as 
$M_1(\tilde m_1/0.1\ {\rm eV})^2<10^{14}$~GeV \cite{ba00,gi04}, 
and hence the values of
$Y_{B-L}$ obtained would not change significantly
 had we adopted smaller  $M_1$ values.
 We start with
vanishing initial values for $Y_N$ and for all the asymmetries, but
notice that for $\tilde m_1>10^{-2}$~eV the results are insensitive to
the initial values.

In the six different temperature regimes we will consider, additional
interactions will enter into equilibrium at each step as the
temperature of the thermal bath decreases:

\bigskip 
\noindent
1) {\it Only gauge and top-Yukawa interactions in equilibrium}
($T> 10^{13}\,$GeV). \\[3pt]
Since in this regime the electroweak sphalerons are out of equilibrium, no
baryon asymmetry is generated during leptogenesis. Moreover, 
since the charged lepton Yukawa interactions are negligible,  the lepton
asymmetry is just in the left-handed degrees of freedom and confined 
in the $\ell=\ell_D$ doublet, yielding $Y_{L}=2\, y_{\ell}\ Y^{eq} =-Y_{B-L}$.
As concerns $y_H$,  although initially equal asymmetries are produced
by the decay of the heavy neutrino in the lepton and in the Higgs
doublets, the Higgs asymmetry is partially transferred into a chiral
asymmetry for the top quarks ($y_t -y_{Q_3}\neq 0$) implying
$y_\ell\neq y_H$. 

\bigskip 

\noindent
2) {\it Strong sphalerons in equilibrium} ($T\sim 10^{13}\,$GeV). \\[3pt]
QCD sphalerons equilibration occurs at higher temperatures than for the
corresponding electroweak processes because of their larger rate
($\Gamma_{QCD} \sim 11 (\alpha_s/\alpha_W)^5\Gamma_{EW}$~\cite{mo97}). 
These processes
are likely to be in equilibrium already at temperatures $T_s\sim 10^{13}$~GeV
\cite{mo97,be03,mo92}) and yield the constraint
\begin{equation}
\sum_i\left(2 \mu_{Q_i}-\mu_{u_i}-\mu_{d_i} \right)=0\,.
\label{qcdsph}
\end{equation}
Direct comparison with the previous case allows us to estimate the
corresponding effects: while the relation $Y_{L}=2\, y_\ell Y^{eq}
=-Y_{B-L}$, implying $c_\ell= 1/2$, holds also for this case, we see
that switching on the QCD sphalerons reduces the Higgs number
asymmetry by a factor of $21/23$. This effect yields a suppression of
the washout that does not exceed 5\%.

%%%%%%%%%%%%%%%%%%%%%%%%%%%%%%%%%%%%%%%%%%%%%%%%%%%%%%%
%%%%%%%%%%%   TABLE %%%%%%%%%%%%%%%%%%%%%%%%%%%%%%%%%%
%%%%%%%%%%%%%%%%%%%%%%%%%%%%%%%%%%%%%%%%%%%%%%%%%%%%%%%
%
\begin{table}[t!] 
\begin{center}
  \renewcommand{\arraystretch}{1.4}
\begin{tabular}
{p{0cm}|>{\centering\small}p{1.4cm}>{\raggedleft\arraybackslash\small}p{2.3cm}>
{\raggedleft%\arraybackslash
\small}c>{\raggedleft\small}c>{\raggedright%\arraybackslash
\small}p{.4cm}c}
%HEADING 
%level 1
\omit & \multicolumn{6}{c}{\bf Equilibrium processes, constraints,
  coefficients and $B-L$ asymmetry} 
\\ \hline\hline 
%%%%%%%%%%%%%%
&&&&& \\[-14pt]
&{\small$T\,$(GeV)}& {\small Equilibrium}  & {\small Constraints} & 
$\hbox{\large \it c}_\ell$ &
$\hbox{\large \it c}_H$&
$\left|\frac{Y_{B-L}}{10^{-5}\epsilon}\right|$ 
 \\[4pt]  \hline

%%%%%%%%%%%%%% I
& $\ \ \gg 10^{13}$ & \quad $h_t$, gauge  & 
$B=\sum_i(2Q_i+u_i+d_i)=0$\hspace{.1cm} 
                              & $\frac{1}{2} $ & $\frac{1}{3}\, $ & 0.6 \\ %[4pt] % \hline
%%%%%%%%%%%%%% II
 &$\  \sim 10^{13}$ &  +\ QCD-Sph  & $\sum_i(2Q_i-u_i-d_i)=0\hspace{1.cm} $
                              & $\frac{1}{2} $ & $\frac{7}{23}\, $ & 0.6  \\ [4pt] % \hline
%%%%%%%%%%%%%% III
\begin{rotate}{90}
$\ B=0$ 
\end{rotate}
&$10^{12\div 13}$ & + \quad  $h_b$,\ $h_\tau \quad $ &
$
\begin{array}{l}
\raise 3pt \hbox{$b=Q_3-H,$}\\[-2pt] 
 \tau=\ell_\tau-H 
\end{array}
\left\{\begin{array}{l}
\!\!\! K_\tau\!=\!0 \\[-4pt] 
\!\!\! K_{\tau}\!=\!1
\end{array}\right.$ &
$\begin{array}{l}\!  
   \frac{1}{2}  \\[-2pt]
   \! \frac{3}{8}
\end{array}$ &  
    $\hspace{-2mm} 
\begin{array}{l}  
\frac{3}{16} \\[-2pt] 
\> \frac{1}{4}
\end{array}$ & 
$\begin{array}{l}  
0.7 \\[-4pt] 
0.8
\end{array}$  
    \\[20pt] \hline 
%%%%%%%%%%%%%%%%%%%%%%%%%%%%%%%%%%%%%%%%%%%%%%%%%%%%%%%%%%% IV
 &&&&& \\[-16pt]
&$10^{11\div 12}$  &  +\quad  EW-Sph & $\sum_i(3Q_i+\ell_i)=0\ \ 
                             
\left\{\begin{array}{l}
\!\!\! K_\tau\!=\!0 \\[-2pt] 
\!\!\! K_{\tau}\!=\!1
\end{array}\right.$ &
    $\begin{array}{l}
\!\! \frac{49}{115}   \\[-2pt]
\!\! \frac{39}{115}
\end{array}$ 
&  
    $\hspace{-2mm} 
    \begin{array}{l}\, 
  \frac{41}{230}   \\[-0pt]
  \, \frac{28}{115}
\end{array}$ 
& 
    $\begin{array}{l} 
  0.8   \\[-2pt]
  0.9
\end{array}$  
    \\[14pt] 
%%%%%%%%%%%%%% 
\begin{rotate}{90}
$ B\neq 0$ 
\end{rotate}
%%%%%%%%%%%%%% V
&$10^{8\div 11}$ & +\ $h_c$, $h_s$, $h_\mu$  
& $ \begin{array}{l}
c\!=\!Q_2\!+\!H, \\[-3pt] 
s\!=\!Q_2\!-\!H,  \\[-3pt] 
\mu\!=\!\ell_\mu\!-\!H
\end{array}
\ \left\{ \begin{array}{l}
\!\!\! K_e=\!1 \\ [-2pt]
\!\!\! K_\tau\!=\!1
\end{array}\right.\ \ $ &                  
$\begin{array}{l}
\!\!   \frac{151}{358}  \\[-2pt]
\!\!  \frac{172}{537}
\end{array}$ &  
$\begin{array}{l}
\! \frac{37}{358} \\[-2pt]
\! \frac{26}{179}
\end{array}$   & 
$\begin{array}{l}
\, 1.0 \\[-2pt]
\, 1.1
\end{array}$ 
   \\[20pt] 
%%%%%%%%%%%%%% VI
& $\ll 10^{8}$ 
& % + $h_u$, $h_d$, $h_e$  & % \multicolumn{2}{l}
$ \begin{array}{l}  
  \hbox{\rm \small
    all Yukawas}\ h_i
 \end{array} $ 
&
$K_e=\!1$
  & 
  $\frac{221}{711}$
& 
$\frac{8}{79}$
&
$1.2$
\\[12pt]
\hline \hline
\end{tabular}
\caption{\baselineskip 12pt 
  The relevant quantities in the different temperature
  regimes. Chemical potentials are labeled here with the same notation
  used for the fields: $\mu_{Q_i}\!=\!Q_i$, $\mu_{\ell_i}\!=\!\ell_i$
  for the $SU(2)$ doublets, $\mu_{u_i}\!=\!u_i$, $\mu_{d_i}\!=\!d_i$,
  $\mu_{e_i}\!=\!e_i$ for the singlets and $\mu_H=H$ for the Higgs.
  The relevant reactions in equilibrium in each regime are given in
  the second column and the constraints imposed on the third. The
  alignment conditions adopted for  the  $K_i$ are indicated. The
  appropriate constraints on the conserved quantities $\Delta_i \!=\!
  B/3 \!-\!  L_i$ should also be imposed.  The values of the
  coefficients  $c_\ell$ and $c_H$ are given respectively in the
  fourth and fifth column while the resulting $B-L$ asymmetry (in
  units of $10^{-5}\times \epsilon$) obtained for $\tilde m_1 =
  0.06\,$ eV and $M_1= 10^{11}\,$GeV is given in the last column.}
\label{tab:results}
\end{center}
\end{table}

%%%%%%%%%%%%%%%%%%%%%%%%%%%%%%%%%%%%%%%%%%%%%%%%%%%%%%%%
%%%%%%%%%%%%%%%%%%%%%%%%%%%%%%%%%%%%%%%%%%%%%%%%%%%%%%%%

\noindent
3) {\it Bottom- and tau-Yukawa interactions in equilibrium}
($10^{12}\ {\rm GeV}\lsim T\lsim 10^{13}\,$GeV). \\[3pt] 
Equilibrium for the bottom and tau Yukawa interactions implies that
the asymmetries in the $SU(2)$ singlet $b$ and $e_\tau$ degrees of
freedom are populated. The corresponding chemical potentials obey the
equilibrium constraints $\mu_b=\mu_{Q_3}-\mu_H$ and
$\mu_\tau=\mu_{\ell_\tau}-\mu_H$.  Possibly $h_b$ and $h_\tau$ Yukawa
interactions enter into equilibrium at a similar temperature as the
electroweak sphalerons \cite{be03}. However, since the rate of the
non-perturbative processes is not well known, we first consider the
possibility of a regime with only gauge, QCD sphalerons and the Yukawa
interactions of the whole third family in equilibrium. This will also
allow us to disentangle by direct comparison with the next case the
new effects induced by electroweak sphalerons.  As regards the flavor
composition of the lepton asymmetry, we distinguish two alignment
cases: first, when the lepton asymmetry is produced in a direction
orthogonal to $\ell_\tau$ ($K_\tau=0$) and second, when it is produced
in the $\ell_\tau$ channel ($K_\tau=1$). 
Since electroweak sphalerons are not yet active, $L_\tau=0$ or
$L_\tau=L$ are conserved quantities in the respective cases. When
$L_\tau=0$, the lepton asymmetry is produced in one of the two
directions orthogonal to $\ell_\tau$ and therefore it does not `leak'
into the $SU(2)$ singlet degrees of freedom, implying that $c_\ell=
1/2$ still holds. In the case when $L=L_\tau$, the washout effects are
somewhat suppressed, since the lepton asymmetry is partially shared
with $e_\tau$ that does not contribute directly to the washout
processes.  Our results for these two cases suggest that the effect on
the final value of $B-L$ associated to the $\tau$ Yukawa interactions
is of the order of 10\%.

Equilibrium for both the top and the bottom quark Yukawa interactions
enforces the constraint $2\mu_{Q_3}-\mu_{u_3}-\mu_{d_3}=0$ and
therefore chemical potentials of the third generation are not
constrained by the QCD sphaleron condition (\ref{qcdsph}). A similar
statement holds for each generation when its quark Yukawa interactions
({\it i.e.} $h_c$ and $h_s$ and, at low enough temperature, also $h_u$
and $h_d$) enter into equilibrium. We conclude that the lower the
temperature that is relevant to leptogenesis, the less important is
the role played by QCD sphaleron effects.

\bigskip

\noindent
4) {\it Electroweak sphalerons in equilibrium }
($10^{11}\ {\rm GeV}\lsim T\lsim 10^{12}\,$GeV). \\[3pt]
The electroweak sphaleron processes take place at a rate per unit volume
$\Gamma/V\propto T^4\alpha_W^5\log(1/\alpha_W)$
\cite{ar98,bodeker98,ar97}, and are expected to be in equilibrium from
temperatures of about $\sim 10^{12}$~GeV, down to the electroweak
scale or below \cite{be03}. Electroweak sphalerons equilibration implies  
\begin{equation}
\label{ewsph}
\sum_i\left( 3\mu_{Q_i}+\mu_{\ell_i} \right)=0\,.
\end{equation}                                
As concerns lepton number, each electroweak sphaleron transition
creates all the doublets of the three generations, implying that
individual lepton flavor numbers are no longer conserved, regardless
of the particular flavor direction along which the doublet $\ell_D$ is
aligned. As concerns baryon number, electroweak sphalerons are the
only source of $B$ violation, implying that baryon number will be
equally distributed among the three families of quarks. In particular,
for the third generation, $B_3=B/3$ is distributed between the
doublets $Q_3$ and the singlets $t$ and $b$.

In Table \ref{tab:results} we give the coefficients $c_\ell$ and $c_H$
for the two aligned cases: (i) $\ell\perp\ell_\tau$ ($K_\tau=0$)
implying  $\Delta_\tau=0$, and (ii) $\ell=\ell_\tau$ ($K_\tau=1$)
implying $\Delta_e=\Delta_\mu=0$. Again we see that the transfer of part
of the lepton asymmetry to a single right handed lepton ($e_\tau$) 
can have a 10\% enhancing effect on the final $B-L$. 

\bigskip
\noindent
5) {\it Second generation Yukawa interactions in equilibrium}
($10^{8}\ {\rm GeV}\lsim T\lsim 10^{11}\,$GeV). \\[3pt] In this regime, the
$h_c$, $h_s$ and $h_\mu$ interactions enter into equilibrium. We
consider two cases of alignment: (i) $\ell=\ell_e$ ($K_e=1$), implying
$\Delta_\tau=\Delta_\mu=0$, and (ii) $\ell\perp\ell_e$.  To ensure a
pure states regime we further assume complete alignment of $\ell$ with
one of the two flavors with Yukawas in equilibrium, for definiteness
$\ell_\tau$, and therefore we have $K_\tau=1$ and
$\Delta_e=\Delta_\mu=0$.  The difference in $c_\ell$ between the two
aligned cases is larger than in the regimes 3 and 4; this, however, is
well compensated by an opposite difference in $c_H$, keeping the
effect on $B-L$ at the same level as in the cases in which just the
third generation Yukawa couplings are in equilibrium.

\bigskip 
\noindent
6) {\it All SM Yukawa interactions in
  equilibrium} ($T\lsim 10^{8}\,$GeV). \\[3pt]
In this regime, since all quark Yukawa interactions are in
equilibrium (actually this only happens for $T<10^6$~GeV), the QCD
sphaleron condition becomes redundant. Hence ignoring the constraint
of eq.~(\ref{qcdsph}), as is usually done in the literature, becomes
fully justified only within this regime. If, however, leptogenesis
takes place at $T\gg 10^6$~GeV, as favored by theoretical
considerations, the constraint implied by the QCD sphalerons is
non-trivial, even if the associated numerical effects are not large. 

Due to the symmetric situation of having all Yukawa interactions in
equilibrium we have just one possible flavor alignment (the other two
possibilities being trivially equivalent).  We take for definiteness
$\ell=\ell_e$ ($K_e=1$) implying $\Delta_\tau=\Delta_\mu=0$.  We see
that in this case $c_\ell$ is reduced by a factor of almost two with
respect to the case in which the spectator processes are neglected
($c_\ell =1/2$) and the final value of $B-L$ is correspondingly
enhanced. The reason for the reduction in $c_\ell$ can be traced
mainly to the fact that a sizable amount of $B$ asymmetry is being
built up at the expense of the $L$ asymmetry, and also a large
fraction of the asymmetry is being transferred to the right handed
degrees of freedom at the same time when inverse decays and washout
processes are active, reducing the effective value of $y_\ell$ that
contributes to drive these processes.

%%%%%%%%%%%%%%%%%%%%
\subsection{Discussion} 
\label{sec:con}

\begin{figure}[ht]
\centerline{\protect\hbox{
\epsfig{file=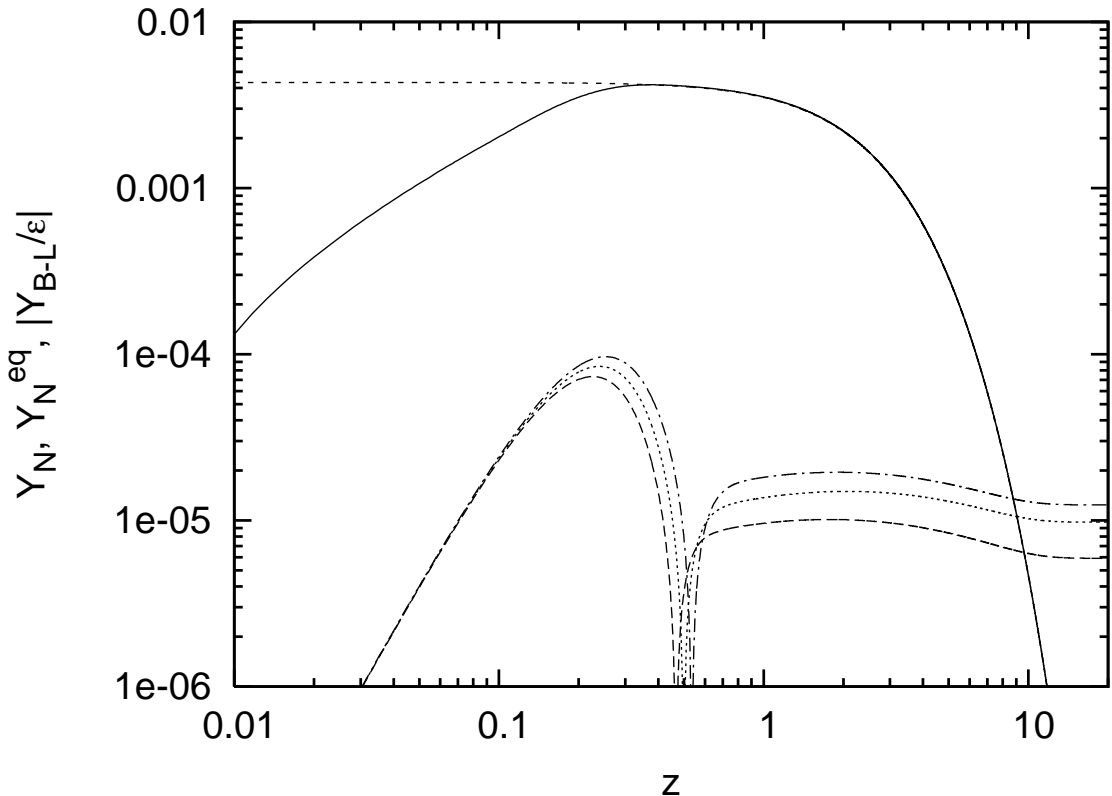 ,width=11cm,angle=270}}}
%\vskip 1.0 truecm
\caption[]{The heavy neutrino density $Y_N$ (solid) and equilibrium
  density $Y_N^{eq}$ (short-dashes), and the $B-L$ asymmetry
  $|Y_{B-L}/\epsilon|$ for three sets of  values for $(c_\ell,c_H)$:
  (i) the long dashed curve corresponds to $c_\ell=1/2$, $c_H=1/3$
  (first row in Table \ref{tab:results}); (ii) the dot-dashed curve
  corresponds to  $c_\ell=221/711$, $c_H=8/79$ (last row in Table
  \ref{tab:results}); (iii) the dotted curve corresponds to
  $c_\ell=-1/2$, $c_H=0$. We take $M_1=10^{11}$~GeV and $\tilde
  m_1=0.06$~eV.}  
\label{fig3}
\end{figure}

The range of final asymmetries presented in Table \ref{tab:results},
that correspond to the cases of flavor alignment, gives a measure of
the possible  impact of the spectator processes (ignoring flavor 
issues) in the different regimes. In Fig.~\ref{fig3} we show the
results of integrating the Boltzmann equations with the two pairs of
extreme values of $c_{\ell,H}$ given in the first and in the last row
of Table~\ref{tab:results}. We also show the results for the
(incorrect) case in which only the asymmetry $y_\ell$ is included in
the washout terms, and all the effects of the spectator processes
discussed in this paper are ignored ($c_\ell=1/2$ and $c_H=0$).  We
learn that when the electroweak sphalerons are not active and flavor
effects are negligible, the Higgs contribution enhances the washout
processes, leading to a smaller final $B-L$ asymmetry. As more and
more spectator processes become fast (compared to the expansion rate
of the Universe), the general trend is towards reducing the value of
the washout coefficients and hence increasing the final value of the
resulting $B-L$ asymmetry. A rough quantitative understanding of our
results can be obtained relying on the fact that the surviving
asymmetry is inversely proportional to the washout rate, as can be
demonstrated along lines similar to those given in Appendix 2 of
ref. \cite{ba00}.  Hence, the relative values for the final $B-L$
asymmetries obtained in the relevant temperature regimes in
Table~\ref{tab:results} can be roughly explained as being inversely
proportional to $c_\ell+c_H$.\footnote{The washout rate
having the strongest impact on the final value of the asymmetry for
$M_1<10^{14}~{\rm GeV}\times({\rm 0.1\, eV}/\tilde m_1)^2$ is the
$\Delta L=2$ on-shell piece of $\gamma_{N_s}$, which has a Boltzmann
suppression factor exp($-z$), similar to the $\Delta L=1$ rates. Hence
the proof given in ref.~\cite{ba00} for the case of $\Delta L=1$
washout dominance and small departure from equilibrium holds also for
the cases considered in Table~\ref{tab:results}.}  The largest value
for $B-L$ given in the table corresponds to the case in which we
assumed all the Yukawa interactions in equilibrium during the
leptogenesis era.  This result is different from the one obtained in
\cite{bu01}, where an order one enhancement of the washout processes
was found for this same case, and hence a smaller final $B-L$
asymmetry.  (In more general non-aligned flavor configurations this
disagreement would be even more pronounced \cite{future}.)  
In \cite{bu01} the washout term involving the leptonic
asymmetry was taken to be proportional to the total asymmetry $Y_L$
rather than just to the asymmetry in the lepton doublet $\ell $, and
we think that this may be the main cause of the discrepancy.

%%%%%%%%%%%%%%%%%%%%%%%%%%% 
\section{Implications for light neutrino masses}  
\label{sec:mbar}
Leptogenesis, besides providing an attractive mechanism to account for
the baryon asymmetry of the Universe, has interesting implications for
low energy observables. In particular, assuming that leptogenesis is
indeed the source of the baryon asymmetry, the observed value of this
quantity then implies a strong upper bound on the absolute scale of
the light neutrino masses. In this section we discuss the implications
of our analysis for this bound. Note that we are concerned here with
the high temperature regime $T\gg 10^{13}\,$GeV for which flavor
considerations are not relevant. Therefore, the simplifying
flavor-related assumptions that we make in this work are fully
justified for the purposes of this section.

The numerical value of the baryon to entropy ratio can be expressed 
as
\begin{equation}
  \label{barasy}
\frac{n_B-\bar n_B}{s}=-1.38\times10^{-3}\ \epsilon\ \eta \ \simeq\
8.7\times 10^{-11}, 
\end{equation}
and the upper bound implied for the mass of the heaviest neutrino 
reads~\cite{Buchmuller:2002jk,Hambye:2003rt,Buchmuller:2003gz,Buchmuller:2004nz}: 
\begin{equation}
m_3\lsim 0.15\ {\rm eV}.
\label{mnulim}
\end{equation}
In eq.~(\ref{barasy}) the washout factor $\eta$ is related to the
 various lepton number violating processes of eq. (\ref{eqyl}) and
 depends on the coefficients $c_\ell$ and $c_H$. Then if the effect of
 the Higgs asymmetry, that in the high temperature regimes contributes
 to the washout reducing the value of $\eta$, is taken into account,
 this could result in strengthening the bound~(\ref{mnulim}) on $m_3$.
 This bound lies in the region of quasi-degenerate light neutrinos,
 that is, $(m_3)_{\rm max}\gg m_{\rm atm}\equiv\sqrt{\Delta m^2_{\rm
 atm}}\sim 0.05\ {\rm eV}$.  Thus, we can use in a self-consistent way
 the approximation
\begin{equation} 
m_1\simeq m_2 \simeq m_3\quad {\rm with}\quad 
m_3^2-m_1^2\simeq\Delta m^2_{\rm atm},
\end{equation} 
and  neglect $\Delta m^2_{\rm sol}=m_2^2-m_1^2\ll \Delta m^2_{\rm atm}$. 

The maximal value of the CP asymmetry $\epsilon$, for quasi-degenerate
light neutrinos (and hierarchical heavy neutrinos), is given by
\cite{Hambye:2003rt,Buchmuller:2003gz,Davidson:2002qv} 
\begin{equation}\label{epsmax}
\epsilon_{\rm max}=\frac{3}{32\pi}\frac{\Delta m^2_{\rm
    atm}}{v^2}\frac{M_1}{m_3}\sqrt{1-\frac{m_1^2}{\tilde m_1^2}}.
\end{equation}
In order to set an upper bound on the neutrino masses, the relation 
$m_3={\rm max}(m_{\rm atm}, \tilde m_1)$  is often adopted
\cite{ba00}. With this plausible ansatz we see that, for $\tilde
m_1>m_{\rm atm}$ one has $\epsilon_{\rm max}\propto M_1/m_3^2$.
(For $\tilde m_1\leq m_{\rm atm}$, one has instead $\epsilon_{\rm max}\propto
M_1/{m_3}$.)  

As concerns the washout factor, the lower bound on the $\tilde m_1$
parameter, $\tilde m_1\geq m_1$, implies that for quasi-degenerate
neutrinos one is in the strong washout regime, defined by (see, for
example, \cite{Buchmuller:2004nz})
\begin{equation} 
\tilde m_1\gg  
\tilde m_1^\ast\equiv \frac{256 g_{\rm SM}v^2}{3 M_{Pl}}\simeq 2.3\times 10^{-3}\
       {\rm eV}, 
\end{equation} 
where $g_{\rm SM}=118$ is an effective number of degrees of freedom for $T\gg
100$~GeV. Within the strong washout regime, we distinguish between two regions:

(i) For $M_1<10^{14}~{\rm GeV}(0.1\ {\rm eV}/\tilde m_1)^2$, $\eta$ is
inversely proportional to the strength of the on-shell washout
rates. More precisely, a fit to $\eta$ valid for $\tilde m_1>\tilde
m_1^\ast$ (small departure from equilibrium) and $M_1<10^{14}~{\rm
GeV}(0.1\ {\rm eV}/\tilde m_1)^2$ gives \cite{ba00}:
\begin{equation}
\eta\simeq\frac{1}{\lambda}\left(\frac{\tilde m_1}{0.55\times 10^{-3}\
    {\rm eV}}\right)^{-1.16} \ \equiv \eta_l.
\label{etafit}
\end{equation}
Notice that we introduced here the factor $\lambda\equiv (c_\ell+c_H)/0.5$
to account for the scaling of the rates.

(ii) For $M_1>10^{14}~{\rm GeV}(0.1\ {\rm eV}/\tilde m_1)^2$,
contributions associated to off-shell $N_\alpha$ exchange become the
dominant washout processes, and give rise to an exponential
suppression of $\eta$, with exponent proportional to the square root
of the  $\Delta L=2$ rates~\cite{ba00}:
\begin{equation}
\eta\simeq {\rm exp}\left[-\frac{\tilde m_1}{\tilde m_1^\ast}
\sqrt{\lambda\frac{M_1}{M_1^\ast}X}\right]\equiv \eta_h.
\label{etamay} 
\end{equation} 
Here  $M_1^\ast\simeq  3.3\times  10^{15}$~GeV  and  $X\geq  1$  is  a
parameter  related  to  the  flavor  structure  of  the  $\Delta  L=2$
off-shell processes,  which can  be taken as  $X\simeq 1$  for $\tilde
m_1\simeq m_3$ (see  \cite{ba00}). Since for $M_1\gg 10^{12}$~GeV
no   leptonic  Yukawa   couplings  are   in  equilibrium   during  the
leptogenesis  era,   flavor alignment  issues in  the Boltzmann
equations can be ignored and the effects of the $\Delta L=2$ rates are
just proportional to $\lambda$.

In the regime in which $\eta\simeq \eta_l$ one has that
$n_B/s|_{max}\propto M_1/m_3^{3.16}$ and hence for a given value of
$M_1$ and upper bound on $m_3$ results.  For increasing values of
$M_1$ the bound gets correspondingly relaxed, until for $M_1 \sim
10^{14}~{\rm GeV}(0.1\ {\rm eV}/\tilde m_1)^2$ we approach the regime
in which $\eta\simeq \eta_h$.  In this regime the maximal CP asymmetry
$\epsilon_{\rm max}$ still increases with $M_1$; however, due to the
exponential suppression of the efficiency factor $\eta_h$, here the
upper bound on $ m_3$ gets  strengthened with increasing $M_1$.  Now,
if one tries to bound $ m_3$ by looking for the value of $M_1$ which
maximizes the product $\epsilon_{\rm max}\eta_h$, one finds
that this arises for values of $M_1$ in the transition region, when
$\eta$ changes from $\eta_l$ to $\eta_h$.  Since the precise bound
depends on the exact way in which $\eta$ interpolates between the two
asymptotic behaviors, a detailed numerical analysis is required for a
reliable estimate of the limit, and in particular to determine the
effect of the Higgs asymmetry on its value.

Nevertheless, in order to get some insight into the possible scaling
behavior of the limit with $\lambda\neq 1$, let us proceed
analytically by adopting the simple interpolation  
\begin{equation} 
\eta\simeq \left(\frac{1}{\eta_l}+\frac{1}{\eta_h}\right)^{-1},
\label{interpol} 
\end{equation} 
that gives a reasonable fit to the detailed numerical results obtained
in \cite{ba00} and \cite{gi04} for $\tilde m_1>\tilde m_1^\ast$.
In general the maximum value of $m_3$ leading to successful
leptogenesis, i.e. to $n_B/s>9\times 10^{-11}$, is obtained by looking
to the parametric curves $ m_3(M_1)$ corresponding to $n_B/s=9\times
10^{-11}$, and requiring that d$m_3/{\rm d}M_1=0$. It is easy to show
that, at fixed $m_3$, the baryon asymmetry is then maximized for a
value of $M_1$ satisfying:
\begin{equation}
\frac{{\rm d\ }\ln\eta_h}{{\rm d}M_1}=-\frac{\eta_h}{\eta M_1}\
\Longrightarrow\ 
M_1\simeq \frac{4}{\lambda}
\ \left(\frac{\xi \tilde m_1^\ast}{ m_3}\right)^2 M_1^\ast \equiv
\bar M_1,
\label{moneo}
\end{equation}
where for convenience we have introduced the factor $\xi\equiv
\eta_h/\eta>1$. For $M_1=\bar M_1$ we have $\eta={\rm
exp}(-2\xi)/\xi$ and since the upper bound on $m_3$ is associated
with values of $\eta\simeq 10^{-3}$, we can expect a typical value
$\xi\simeq 3$, that yields $\bar M_1\simeq 6\times 10^{13} \lambda^{-1}
(0.1\ {\rm eV}/ m_3)^2$~GeV.
Defining $\bar\eta\equiv \eta(\tilde m_1=\bar m_3,M_1=\bar M_1)$, 
the maximum value of $m_3$ that results then is
\begin{equation}
  \bar m_3\lsim 0.19\ {\rm eV}\
  \left(\frac{\bar\eta \xi^2}{\lambda\ 10^{-2}}\right)^{1/4},
\label{barm}
\end{equation}
that is in reasonable agreement with the results of dedicated
numerical analyses. Note, however, that the scaling behavior of this
bound under a change in $\lambda$ can be different from what is
implied by the explicit dependence $\lambda^{-1/4}$, since also the
parameters $\bar\eta$ and $\xi$ depend, in general, on $\lambda$.
Finding the real $\lambda$-dependence is  a non-trivial problem.   
We can study this behavior by performing an infinitesimal transformation
$\lambda\to (1+\epsilon)\lambda$, and finding out how the relevant
quantities $X= \bar m_3,\ \bar M_1,\ \eta ,\ \eta_h$ and $\eta_l $
scale under this transformation:
\begin{equation}
 X \propto \lambda^{n_X}\,, \qquad {\rm with} \qquad      n_X =
 \frac{{\rm d} \ln X}{{\rm d} \epsilon}.
\end{equation}
Relating the exponents $n_X$ of the different quantities, it is then
 possible to show that for the particular interpolation we have
 adopted in eq.~(\ref{interpol}) $\bar m_3\propto \lambda^{-0.4} $ is
 obtained. Note however, that the fine details of the transition
 between $\eta_l$ and $\eta_h$ are important for a precise
 determination of the scaling exponent. For example, for the more
 general class of interpolating functions
 $\eta=(\eta_l^{-a}+\eta_h^{-a})^{-1/a}$ one would find
 $n_{m_3}=-0.25$ for $a=0$ and $n_{m_3}=-0.8$ for $a=\infty$. Still,
 in spite of this uncertainty that is intrinsic to the analytical
 approach, the final numerical results for the strengthening of the
 neutrino mass limit with increasing values of $\lambda$ do not differ
 too much.  In particular, in the regime corresponding to $M_1>
 10^{13}$~GeV which is relevant for the neutrino mass bound we have
 $c_\ell+c_H=5/6$ that corresponds to $\lambda=5/3$. Hence the
 bound on $m_3$ will be smaller than what obtained assuming
 $\lambda=1$ by a factor $(5/3)^{n_{m_3}}$,
 whose likely range is in between 0.66 and 0.88. We can conclude that
by taking properly into account $y_H$ in the Boltzmann equations, a
bound on $m_3$ stronger by something of the order $\sim 20$\% could 
be obtained.

To summarize, we have considered the combined effects of the spectator
processes -- Yukawa, strong- and electroweak-sphaleron interactions --
on the $B-L$ asymmetry generated by leptogenesis. The effects range
between reducing the final asymmetry by order 40\%, if the lepton
asymmetry is generated at temperatures higher than $10^{13}$ GeV, to
enhancing it by order 20\%, if the relevant temperature is well below
$10^8$ GeV. (As will be discussed in \cite{future}, when misalignment
in the lepton doublet flavor space between the combination to which
$N_1$ decays and the direction defined by fast Yukawa interactions
occurs, qualitatively different and much stronger effects can arise.)
Spectator processes strengthen the leptogenesis bound on the light
neutrino mass scale by order 20\%.

\bigskip

%%%%%%%%%%%%%%%%%%%%%%%%%%
\section*{Acknowledgments}
Work supported in part by ANPCyT and Fundaci\'on Antorchas, Argentina, by
Colciencias in Colombia under contract 1115-05-13809, and by the Italian
Istituto Nazionale di Fisica Nucleare (INFN).
E.N. acknowledges conversations with Jorge Iv\'an Nore\~na. 
The work of Y.N. is supported by the Israel Science Foundation
founded by the Israel Academy of Sciences and Humanities, by EEC RTN
contract HPRN-CT-00292-2002, by the Minerva Foundation (M\"unchen),
by the United States-Israel Binational Science Foundation (BSF),
Jerusalem, Israel and by the G.I.F., the
German--Israeli Foundation for Scientific Research and Development.

%%%%%%%%%%%%%%%%%%%%%%%%%%%%


\begin{thebibliography}{99}

\bibitem{fu86} M.~Fukugita and T.~Yanagida,
  %``Baryogenesis Without Grand Unification,''
  Phys.\ Lett.\ B {\bf 174}, 45 (1986).
  %%CITATION = PHLTA,B174,45;%%

\bibitem{lu92} M.~A.~Luty,
  %``Baryogenesis via leptogenesis,''
  Phys.\ Rev.\ D {\bf 45}, 455 (1992).
  %%CITATION = PHRVA,D45,455;%%

\bibitem{ba00}
  R.~Barbieri, P.~Creminelli, A.~Strumia and N.~Tetradis,
  %``Baryogenesis through leptogenesis,''
  Nucl.\ Phys.\ B {\bf 575}, 61 (2000)
  [arXiv:hep-ph/9911315] Version 4.
  %%CITATION = HEP-PH 9911315;%%

\bibitem{future}
  E. Nardi, Y. Nir, J. Racker and E. Roulet, work in progress.
  
\bibitem{bu01} W.~Buchmuller and M.~Plumacher,
  %``Spectator processes and baryogenesis,''
  Phys.\ Lett.\ B {\bf 511}, 74 (2001)
  [arXiv:hep-ph/0104189].
  %%CITATION = HEP-PH 0104189;%%

\bibitem{pi05} A. Pilaftsis and T. Underwood,
Phys. Rev. D{\bf 72} (2005) 113001,
[arXiv:hep-ph/0506107].
  
\bibitem{gi04} G.~F.~Giudice {\it et al.},
  %A.~Notari, M.~Raidal, A.~Riotto and A.~Strumia,
  %``Towards a complete theory of thermal leptogenesis in the SM and MSSM,''
  Nucl.\ Phys.\ B {\bf 685}, 89 (2004)
  [arXiv:hep-ph/0310123].
  %%CITATION = HEP-PH 0310123;%%

\bibitem{pi04} A.~Pilaftsis and T.~E.~J.~Underwood,
  %``Resonant leptogenesis,''
  Nucl.\ Phys.\ B {\bf 692}, 303 (2004)
  [arXiv:hep-ph/0309342].
  %%CITATION = HEP-PH 0309342;%%

\bibitem{co96} L.~Covi, E.~Roulet and F.~Vissani,
  %``CP violating decays in leptogenesis scenarios,''
  Phys.\ Lett.\ B {\bf 384}, 169 (1996)
  [arXiv:hep-ph/9605319].
  %%CITATION = HEP-PH 9605319;%%

\bibitem{ha90} J.~A.~Harvey and M.~S.~Turner,
  %``Cosmological Baryon And Lepton Number In The Presence Of Electroweak
  %Fermion Number Violation,''
  Phys.\ Rev.\ D {\bf 42}, 3344 (1990).
  %%CITATION = PHRVA,D42,3344;%%

\bibitem{la00} M.~Laine and M.~E.~Shaposhnikov,
  %``A remark on sphaleron erasure of baryon asymmetry,''
  Phys.\ Rev.\ D {\bf 61}, 117302 (2000)
  [arXiv:hep-ph/9911473].
  %%CITATION = HEP-PH 9911473;%%

\bibitem{mo97} G.~D.~Moore,
  %``Computing the strong sphaleron rate,''
  Phys.\ Lett.\ B {\bf 412}, 359 (1997)
  [arXiv:hep-ph/9705248].
  %%CITATION = HEP-PH 9705248;%%

\bibitem{be03} L.~Bento,
  %``Sphaleron relaxation temperatures,''
  JCAP {\bf 0311}, 002 (2003)
  [arXiv:hep-ph/0304263].
  %%CITATION = HEP-PH 0304263;%%

\bibitem{mo92} R.~N.~Mohapatra and X.~m.~Zhang,
  %``QCD sphalerons at high temperature and baryogenesis at electroweak scale,''
  Phys.\ Rev.\ D {\bf 45}, 2699 (1992).
  %%CITATION = PHRVA,D45,2699;%%

\bibitem{ar98} P.~Arnold, D.~T.~Son and L.~G.~Yaffe,
  %``Hot B violation, color conductivity, and log(1/alpha) effects,''
  Phys.\ Rev.\ D {\bf 59}, 105020 (1999)
  [arXiv:hep-ph/9810216].
  %%CITATION = HEP-PH 9810216;%%

\bibitem{bodeker98} D. B\"odeker,
  %``On the effective dynamics of soft non-abelian gauge fields at finite
  %temperature,''
  Phys.\ Lett.\ B {\bf 426}, 351 (1998)
  [arXiv:hep-ph/9801430].
  %%CITATION = HEP-PH 9801430;%%

\bibitem{ar97} P.~Arnold, D.~Son and L.~G.~Yaffe,
  %``The hot baryon violation rate is O(alpha(w)**5 T**4),''
  Phys.\ Rev.\ D {\bf 55}, 6264 (1997)
  [arXiv:hep-ph/9609481].
  %%CITATION = HEP-PH 9609481;%%

%\cite{Buchmuller:2002jk}
\bibitem{Buchmuller:2002jk}
  W.~Buchmuller, P.~Di Bari and M.~Plumacher,
  %``A bound on neutrino masses from baryogenesis,''
  Phys.\ Lett.\ B {\bf 547}, 128 (2002)
  [arXiv:hep-ph/0209301].
  %%CITATION = HEP-PH 0209301;%%
  
%\cite{Hambye:2003rt}
\bibitem{Hambye:2003rt}
  T.~Hambye {\it et al.},
  %Y.~Lin, A.~Notari, M.~Papucci and A.~Strumia,
  %``Constraints on neutrino masses from leptogenesis models,''
  Nucl.\ Phys.\ B {\bf 695}, 169 (2004)
  [arXiv:hep-ph/0312203].
  %%CITATION = HEP-PH 0312203;%%

%\cite{Buchmuller:2003gz}
\bibitem{Buchmuller:2003gz}% may be affected by a change in $\lambda$, 

  W.~Buchmuller, P.~Di Bari and M.~Plumacher,
  %``The neutrino mass window for baryogenesis,''
  Nucl.\ Phys.\ B {\bf 665}, 445 (2003)
  [arXiv:hep-ph/0302092].
  %%CITATION = HEP-PH 0302092;%%

%\cite{Buchmuller:2004nz}
\bibitem{Buchmuller:2004nz}
  W.~Buchmuller, P.~Di Bari and M.~Plumacher,
  %``Leptogenesis for pedestrians,''
  Annals Phys.\  {\bf 315}, 305 (2005)
  [arXiv:hep-ph/0401240].
  %%CITATION = HEP-PH 0401240;%%
  
%\cite{Davidson:2002qv}
\bibitem{Davidson:2002qv}
  S.~Davidson and A.~Ibarra,
  %``A lower bound on the right-handed neutrino mass from leptogenesis,''
  Phys.\ Lett.\ B {\bf 535}, 25 (2002)
  [arXiv:hep-ph/0202239].
  %%CITATION = HEP-PH 0202239;%%

\end{thebibliography}
\end{document}